\documentclass[prl,twocolumn,superscriptaddress]{revtex4-1}
\usepackage{epsfig}
\usepackage{amssymb}
\usepackage{amsmath}
\usepackage{amsfonts}
\usepackage{graphicx}
\usepackage{bm}
\usepackage{multirow}
\usepackage{epstopdf}
\usepackage{hyperref}
\usepackage[normalem]{ulem}
\usepackage[utf8]{inputenc}

\newcommand {\be} {\begin{equation}}
\newcommand {\ee} {\end{equation}}
\newcommand {\bea} {\begin{eqnarray}}
\newcommand {\eea} {\end{eqnarray}}

\begin{document}
\preprint{APS/123-QED}
\title{
Magnetic-field-driven redistribution between extended and localized electronic states in high-mobility Si MOSFETs at low temperatures
}
\author{V.~M.~Pudalov}
\affiliation{
V.~L.~Ginzburg Research Center for High Temperature Superconductivity and Quantum Materials,
P.~N.~Lebedev Physical Institute, 119991 Moscow, Russia}
\affiliation{
 HSE University, 101000 Moscow, Russia
 }
\author{M.~E.~Gershenson}
\affiliation{
Serin Physics Lab, Rutgers University, Piscataway NJ08854, USA
}

\begin{abstract}
In the study of oscillatory electron transport in high-mobility Si MOSFETs at low temperatures we observed two correlated effects in weak in-plane magnetic fields: a steep decrease of the magnetic susceptibility $\chi^*(H)$ and an increase of the concentration of mobile carriers $n(H)$. We suggest a phenomenological model of the magnetic-field-driven redistribution between the extended and localized electronic states that qualitatively explains both effects. We argue that the redistribution is mainly caused by magnetization of the  large-spin $S\approx 2$ localized states with energies close to the Fermi energy $E_F$,
coexisting with the majority Fermi liquid state. Our findings also resolve a long-standing disagreement between the experimental data on $\chi^*$ obtained in weak ($H\sim k_BT/\mu_B$) and strong ($H\sim  E_F/g\mu_B$) magnetic fields.
\end{abstract}

\maketitle

\section{Introduction}
Dilute two-dimensional systems of electrons represent a very fruitful
playground  for exploration of the physics of strongly interacting charged fermions.
The conventional approach, i.e. the Landau theory of Fermi liquids (FL),
treats the system of interacting electrons as a gas of quasi-electrons
whose parameters are renormalized by interactions \cite{landau, pines}. It is, however, questionable
whether this description remains valid for a two dimensional (2D) system when the
electron-electron interaction energy $E_{ee}$
greatly exceeds the Fermi energy  $E_F$ \cite{varma_02}, i.e. at $r_s\gg 1$ \cite{note}.
Different approaches to this problem gave birth to a plethora of theoretical suggestions
\cite{spivak, dobro_97, chakra_98, chakra_99, khodel_PRB_2020} for the ground states of strongly
 interacting electron systems.

Various correlated systems close to the Mott transition exhibit a tendency
to phase separate in insulator and metallic phases with different densities \cite{castellani_PRB_2001}. This
tendency is suppressed by the long-range Coulomb interaction and by the gate screening (for the 2D gated systems) which favors uniform phases.
Spatial phase separation and the emergence of an inhomogeneous state also often occurs in the vicinity of a phase transition, e.g. between
 superconducting, normal, and magnetically ordered phases \cite{kornilov_PRB_2004, gerasimenko_PRB_2014}.
 More specifically, for a two-dimensional hole system, using the local compressibility measurements Ilani et al. \cite{ilani_Science_2001} observed the emergence of an inhomogeneous state
 in the vicinity of the metal-to-insulator transition.

 The anomalous spin magnetization observed in the in-plane magnetic fields for the strongly correlated system in thermodynamic magnetization \cite{teneh_PRL_2012} and anomalous magnetotransport \cite{morgun_PRB_2016, pudalov_JOSC_2017}    were interpreted as a transition of a dilute 2D system into the two-phase state.
In the two-phase state, the large-spin collective localized states, the so-called spin droplets (SD), emerge and coexist with FL of mobile electrons. The total spin of an individual droplet,  $S=2$ \cite{teneh_PRL_2012}, is almost independent of the carrier density and temperature, whereas the  number of  droplets per unit area  strongly depends on the temperature and the electron density \cite{morgun_PRB_2016}. The idea of the large-spin droplets has received theoretical support \cite{sushkov_PRB_2013}.

The disorder enhanced magnetism in the ground state for restricted geometries has been considered in Refs.~\cite{eisenberg_PRB_1999, brouwer_PRB_1999, kurland_PRB_2000}.  Formation of local regions with nonzero spin density in a disordered 2D system close to the Stoner instability was predicted in Ref.~\cite{narozhny_PRB_2000}. On the other hand, the two-phase state was suggested to emerge from a pure 2D Fermi liquid on the verge of the Wigner crystallization \cite{spivak}.   Electron and hole puddles,
 coupled by tunneling to the edge states,  are believed to be a common feature of 2D topological insulators \cite{glazman_PRL_2013, glazman_PRB_2014}.  Recently, the mesoscopic Stoner instability  in open quantum dots,  tunnel-coupled  to  external fermionic reservoir, was studied theoretically in Ref.~\cite{burmi_PRL_2020}.
 On the experimental side,  for mesoscopic size samples a phase-inhomogeneous state was observed in Ref.~\cite{cobden_PRL_1999}; signatures of spin polarization for a confined few-electrons system
reported in Refs.~\cite{ghosh_PRL_2004, rogge_PRL_2010}
were considered as evidence of interaction-induced collective spin polarization transition.

In the current paper we
report evidence for the presence of inhomogeneous two-phase state over a wide range of carrier concentrations on the ``metallic'' side of the so-called  ``2D metal-insulator'' transition in macroscopic 2D electron system in (100) Si-MOS structures.
Our detailed measurements of the Shubnikov-de Haas (ShdH) oscillations in vector magnetic fields
revealed steep variations of (i) the renormalized spin susceptibility $\chi^*(H)$ and (ii) the mobile carrier density $n_{\rm SdH}(H)$, driven by a weak in-plane field $H_\parallel$.
The unexpected nonmonotonic variations
of the susceptibility ($\sim 10-20\%$) and 2D carrier density ($\sim 3\%$)  correlate with each other  and with thermodynamic spin magnetization of electrons. These correlations
allowed us to identify the origin of these effects.

We associate the density variation with the magnetic field driven redistribution of electrons between two
phases of different electron states. The majority phase is the 2D Fermi liquid of mobile electrons, for which the local carrier density can be determined from the SdH oscillations. The  minority phase, we believe,  consists of  the collective localized spin droplets \cite{teneh_PRL_2012} which are imbedded in the 2D Fermi liquid.

We show that in the studied two-phase system the carrier redistribution  between the two states is driven by the magnetization of the large-spin SD states in the in-plane magnetic field $H_\parallel$.
We tested this idea by considering
a phenomenological thermodynamic model  of the two-phase state and linked the observed changes in the nonlinear SD magnetization with  changes in the FL density $n_{\rm SdH}(H_\parallel)$ and susceptibility $\chi^*(H_\parallel)$.
 The observed
 $\chi^*(H_\parallel)$ dependence  indicates that a widely employed technique of extracting the
2D spin susceptibility from  measurements of the magnetoresistance saturation  \cite{vitkalov_MR, SKD_MR} may provide inaccurate results not only when measurements are performed in the strong fields of the order of $E_F/g\mu_B$ (as noted in Ref.~\cite{zhang_PRL_2006}), but  already in a much weaker field of the order of
temperature.

\section{Experimental}
The ac (13\,Hz) transport  measurements were performed with two
(100)~Si-MOS samples from different wafers Si6-14 and Si3-10 with the peak mobility 2.4\,m$^2$/Vs and
 3.2\,m$^2$/Vs, respectively  (at $T=0.3$\,K). For the resistivity $\rho_{xx}$ measurements we used
 the vector  magnetic field technique
 with two independent superconducting coils. Typical examples of the SdH oscillations at different
in-plane fields and their fitting may be found in Ref.~\cite{gm, pudalov_PRB_2018, pudalov-spinless}.
The oscillatory component $\delta \rho_{xx}$ was shown  earlier to be  well fitted with
conventional Lifshits-Kosevich formula \cite{SdH, pudalov-spinless,  pudalov_PRB_2018}; this enables accurate extraction of $\chi^*$ and $n_{\rm SdH}$. In particular, $\chi^*$ values have been determined
from the oscillation beating with an accuracy of   $\sim (1 - 2)\%$.

The results were available within the temperature range $T<0.5K$, in which
the ShdH oscillations  in weak $H_\perp$ fields \cite{gm} are not damped by temperature.
The new  data reported here coincide in the  $H_\parallel \rightarrow 0$ limit
  with the $\chi^*(0)$ values reported earlier in Ref.~\cite{gm, klimov_PRB_2008}.

\section{Results}
\begin{figure}[h]
\includegraphics[width=200pt]{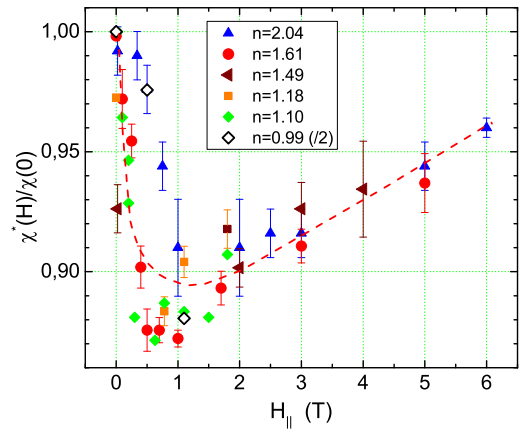}
\caption{
Summary of $\chi^*(H_\parallel)/\chi^*(0)$ data versus $H_\parallel$  for both samples and for several densities.
For the lowest density $n=0.99$, the $\chi^*(H)/\chi^*(0)$ variations are scaled down by two times.
The density is indicated in units of $10^{11}$cm$^{-2}$,  $T=0.1$\,K. The dashed line is a guide to the eye.}
\label{fig1}
\end{figure}

Figure \ref{fig1}  shows  an unexpected nonmonotonic dependence of $\chi^*$ on
 the in-plane field. This dependence is reproducible for  both samples studied.
 As the density increases from $n=0.99$ to  $10\times 10^{11}$cm$^{-2}$, the $\delta\chi^*(H)/\chi^*(0)$ variations decrease from $\sim 25\%$  to $\sim (10-6)\% $.
The characteristic field of the  $\chi^*(H)$-minimum, $H_\parallel \sim 1$\,T for $n=(1.1-2)\times 10^{11}$cm$^{-2}$,  is much weaker than the field of complete spin polarization of the 2D system, $H_p=2E_F/g^*\mu_B$ ($\sim 20$T for $n=2\times 10^{11}$cm$^{-2}$). We note that in a homogeneous single-phase FL-system $H_p$ is the only characteristic field.

\begin{figure}
\includegraphics[width=200pt]{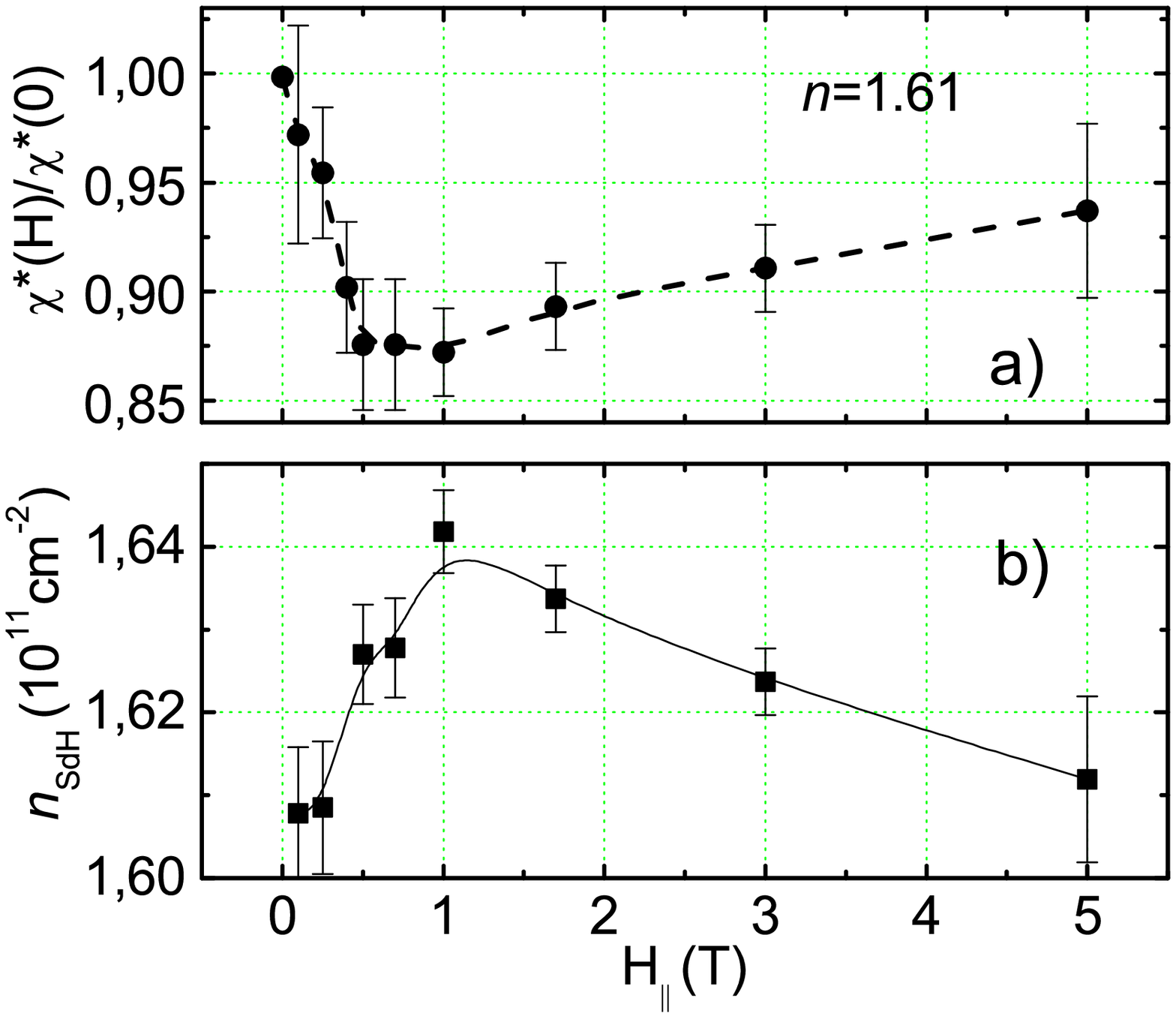}
\hspace*{-0.3cm}
\includegraphics[width=200pt]{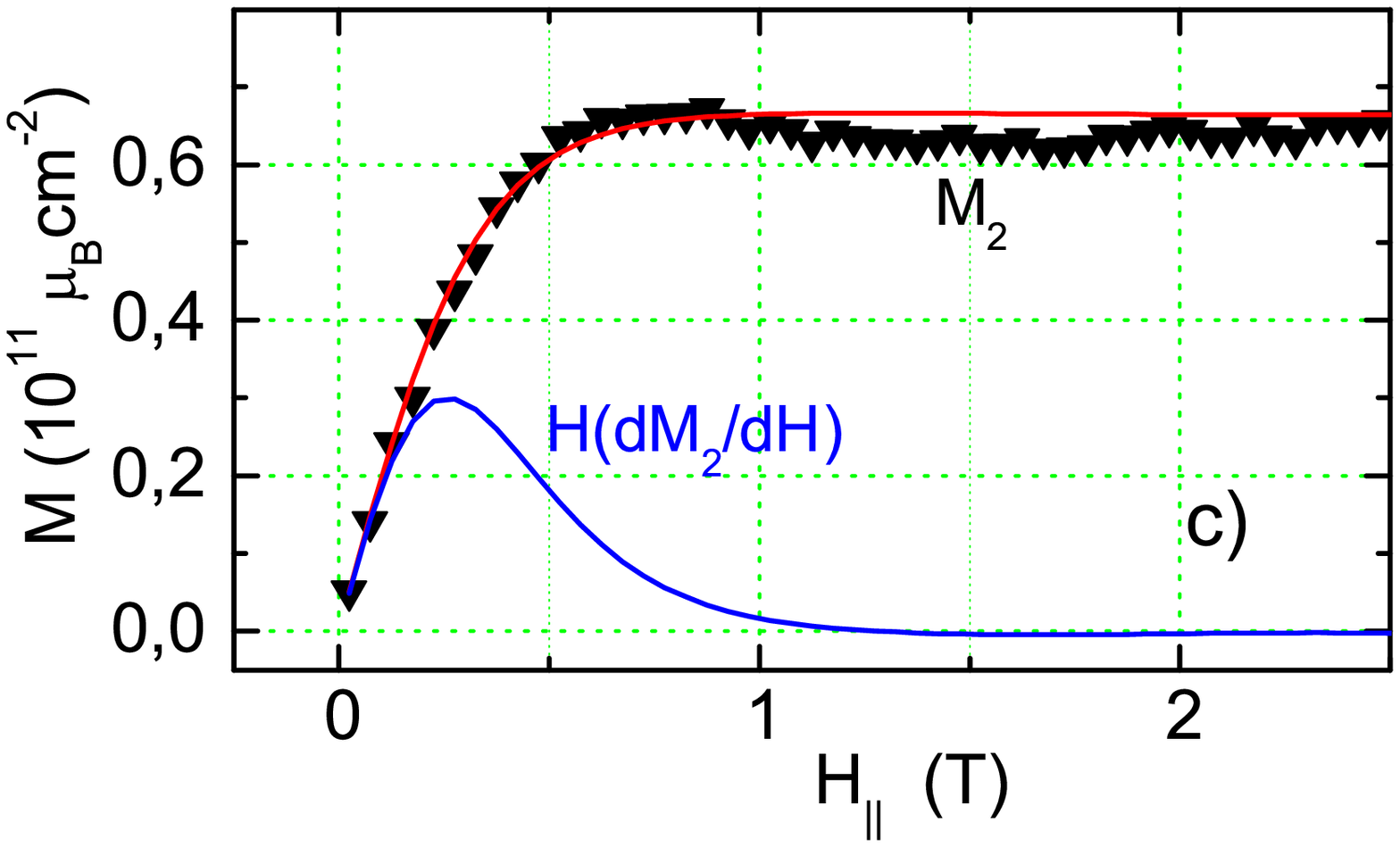}
\caption{Correlation between  the in-plane field dependence of (a) $\chi^*(H)/\chi^*(0)$,
(b) density $n_{\rm SdH}$, and
(c) spin magnetization $M(H_\parallel)$ (reproduced from \cite{pudalov_PRB_2018}). Red curve  shows $\tanh(\mu_B H/k_B T)$-fitting of the experimental $M(H)$ data, blue curve shows $H\partial M/\partial H$ calculated from the fitting curve. The zero-field density  $n_0=1.61\times 10^{11}$cm$^{-2}$ for (a) and (b), and $1.4\times 10^{11}$cm$^{-2}$ for (c). Temperature $T=0.1$\,K for (a) and (b), and 1.7\,K for (c).
}
\label{fig2}
\end{figure}

The hint for explanation of the puzzling $\chi^*(H_\parallel)$ field dependence is provided by the observed remarkable correlation between the spin susceptibility minimum and the maximum of the mobile carrier density, determined from  the  SdH oscillation frequency
in weak tilted fields (see Fig.~\ref{fig2}).
Though the density  changes $\delta n_{\rm SdH}/n_{\rm SdH}$ are small, $\sim 2\%$, they are  reliably determined for various densities.
 The correlation has been observed over the whole studied  range of densities.
Since the total charge of the gated structure is conserved, the observed dependence  $\delta n_{\rm SdH}(H)$ provides evidence for the field-induced redistribution between the localized and extended electronic states.

Both sharp   field dependences of the spin susceptibility  $\chi^*(H)/\chi^*(0)$  and  the    density of mobile carriers $n_{\rm SdH}(H)$
have been measured simultaneously in the same experiment, from beating pattern of SdH oscillations in vector fields, and are in remarkable correlation with each other.
 Such ``generic'' $V$-  and $\Lambda$- shape dependences have been observed
 over the  range of densities $(1.1-2.1)\times 10^{11}$cm$^{-2}$ (referred to as ``intermediate'' densities); they  are qualitatively similar for the two  studied samples.

Below we describe  more complex behavior of $\chi^*(H)$ and $n_{\rm SdH}(H)$ over the extended range of densities.

\subsection{Low densities}
\begin{figure}
\includegraphics[width=240pt]{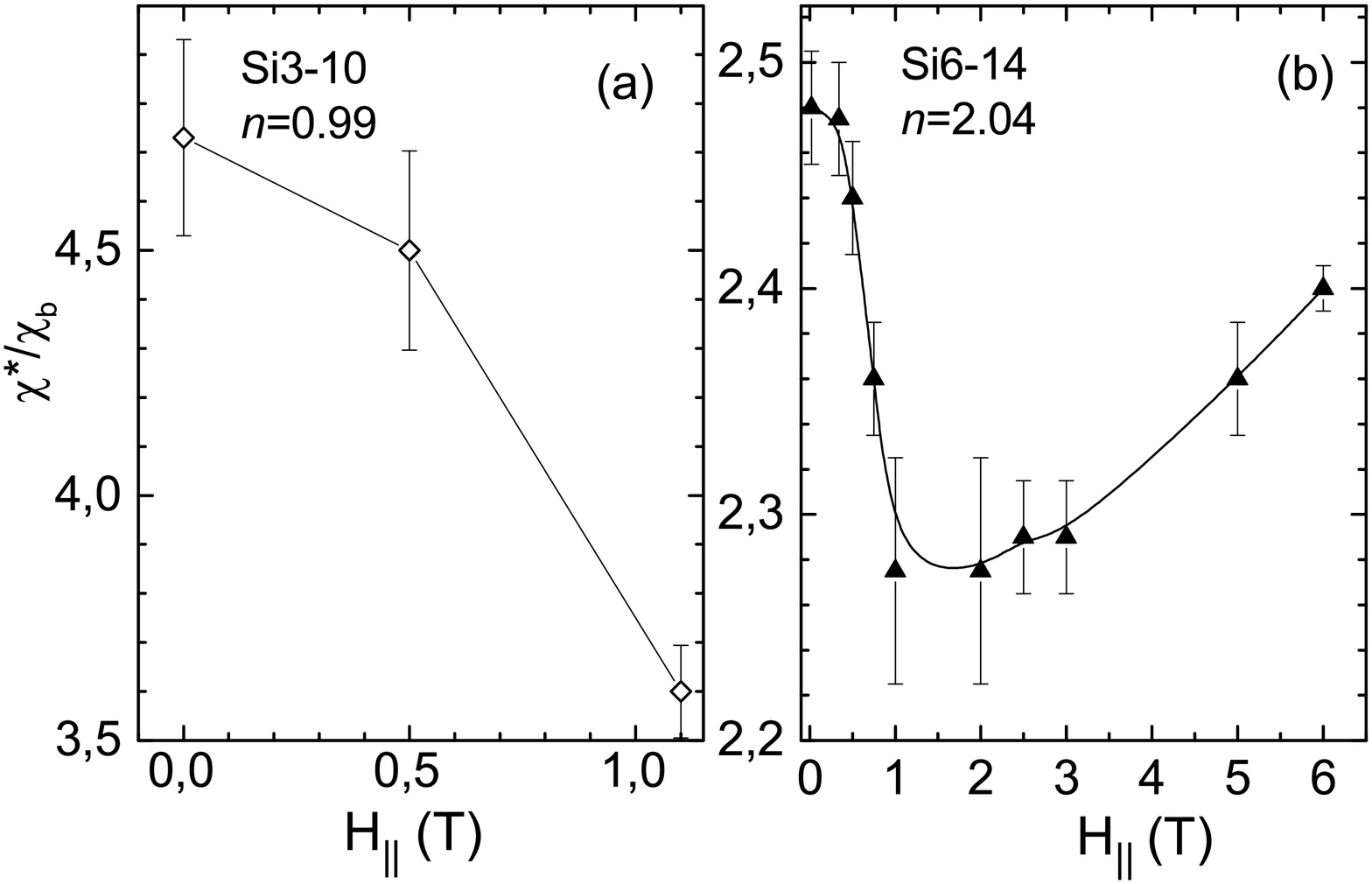}
\includegraphics[width=120pt]{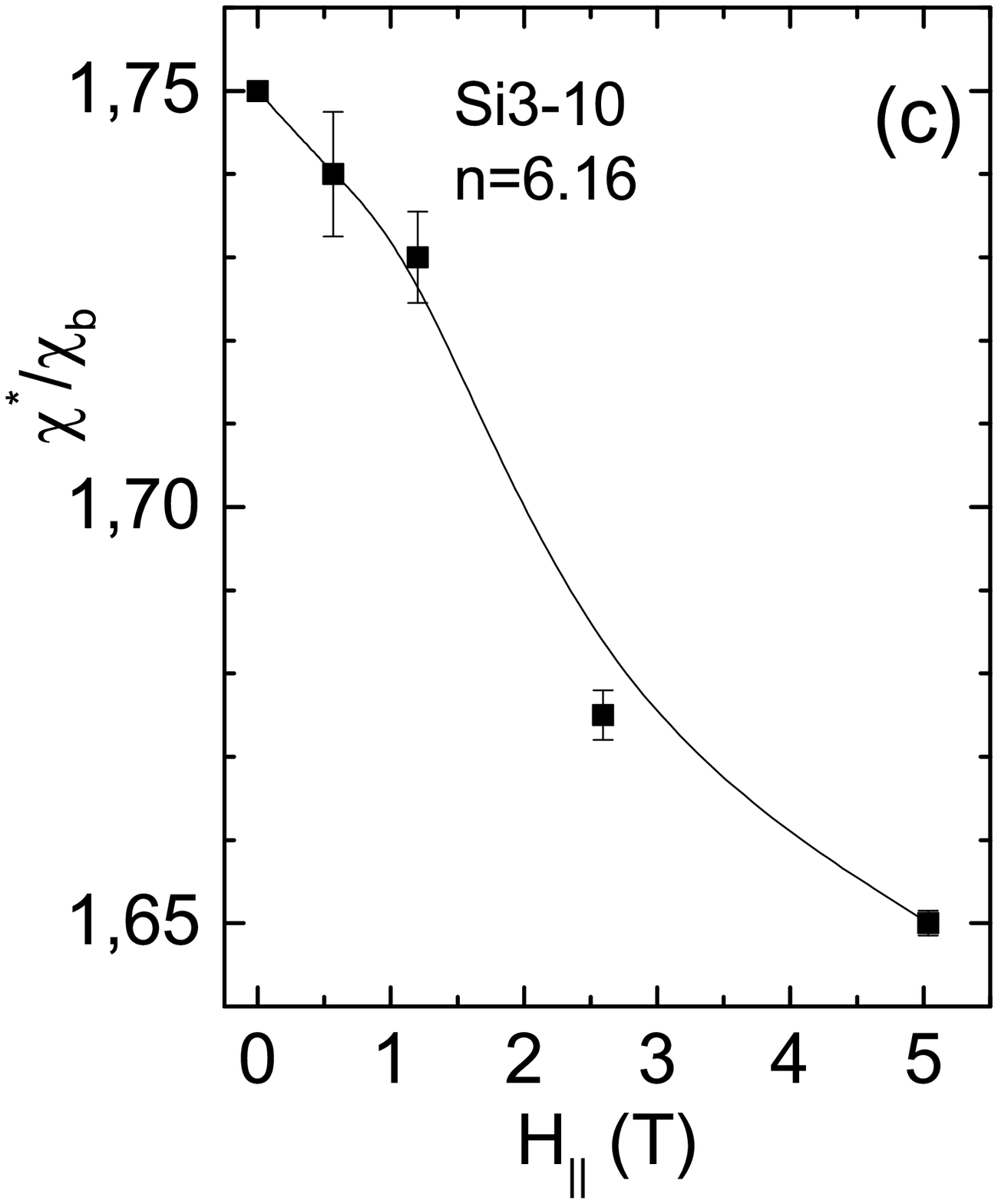}
\includegraphics[width=120pt]{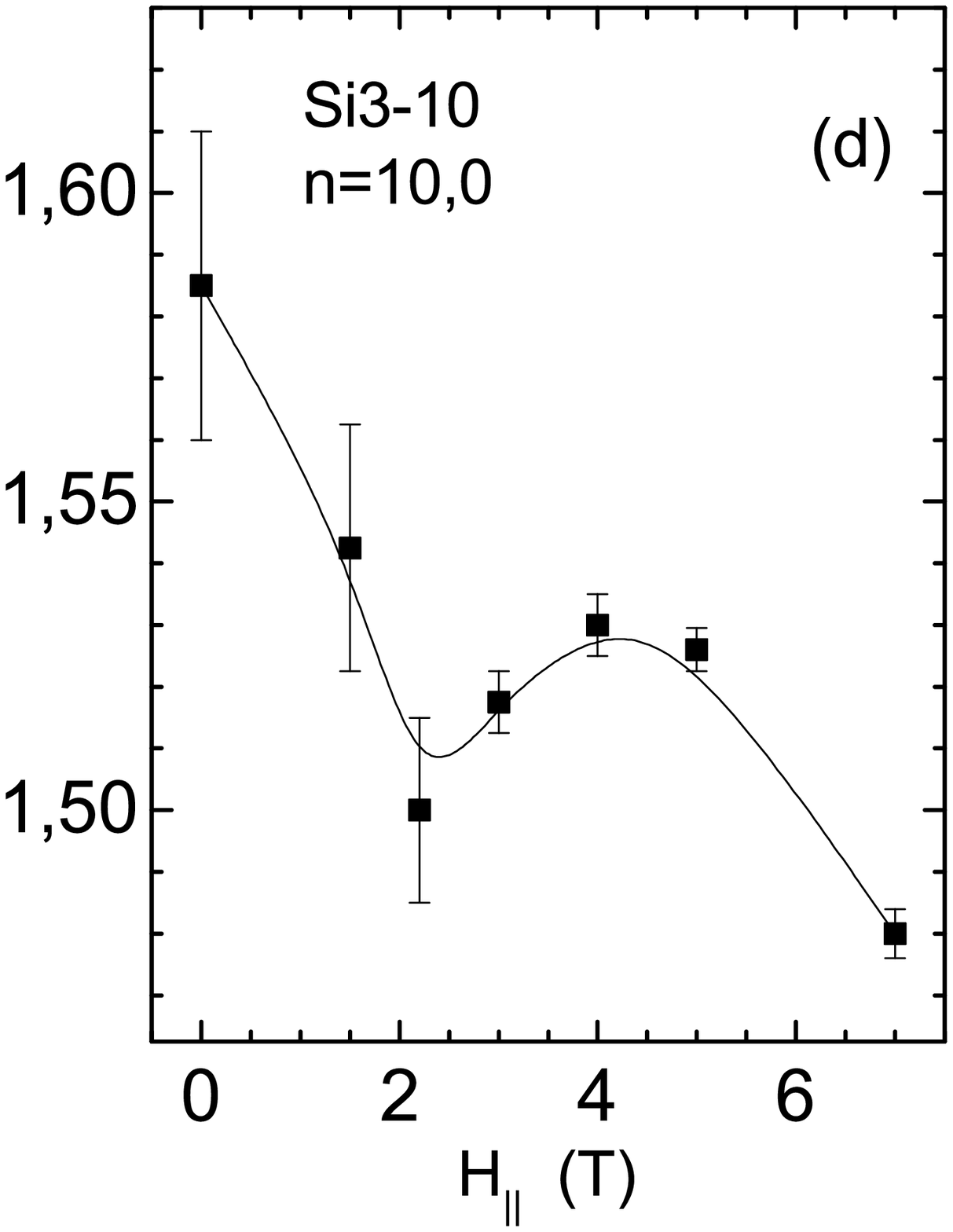}
\caption{Evolution of the  $\chi^*(H)/\chi_b$ dependences
with density, from the lowest density $0.99\times 10^{11}$cm$^{-2}$  (a) to higher densities  $2\times 10^{11}$cm$^{-2}$ (b),  $6.16\times 10^{11}$cm$^{-2}$ (c), and at the highest studied density $10.0\times 10^{11}$cm$^{-2}$ (d).
}
\label{fig3}
\end{figure}

For the lowest density  $n=0.99\times 10^{11}$cm$^{-2}$ the drop  $\delta\chi^*$ with  $H_\parallel$ is surprisingly prominent ($25\%$), as Fig.~\ref{fig3}a shows. At such low densities on the verge of the transition to fully localized state, the $n_{\rm SdH}(H)$ variation could not be measured and variations of $\chi^*(H)$ could not be traced to higher field, because application of an in-plane field caused complete localization of the 2D system  \cite{pud-MR_JETPL_1997, simonian-MR_PRL_1997, pud_physicaB_1998, krav-tilted_PRB_1998}.

\subsection{High densities}
\begin{figure}[h]
\includegraphics[width=195pt]{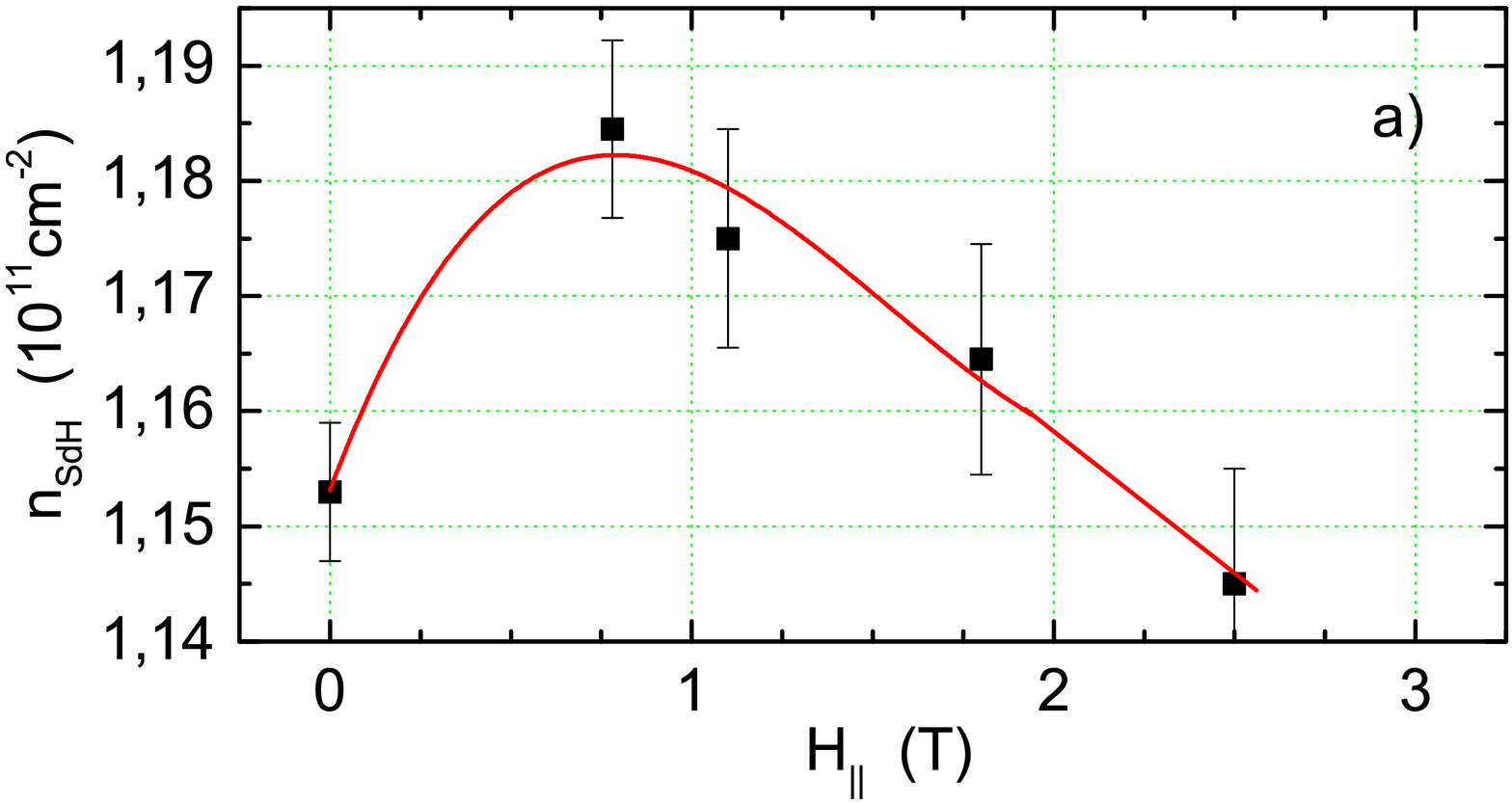}
\includegraphics[width=195pt]{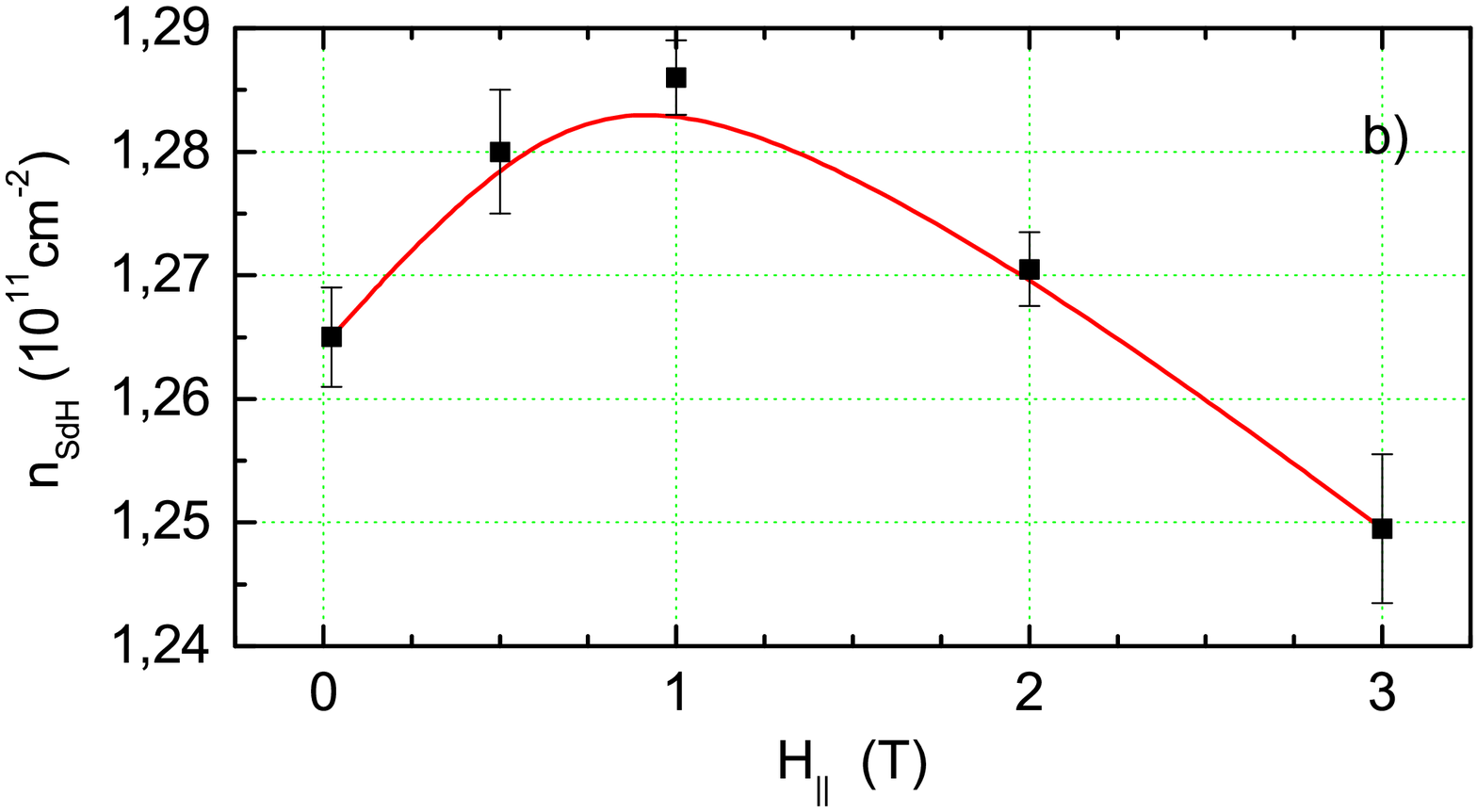}
\includegraphics[width=195pt]{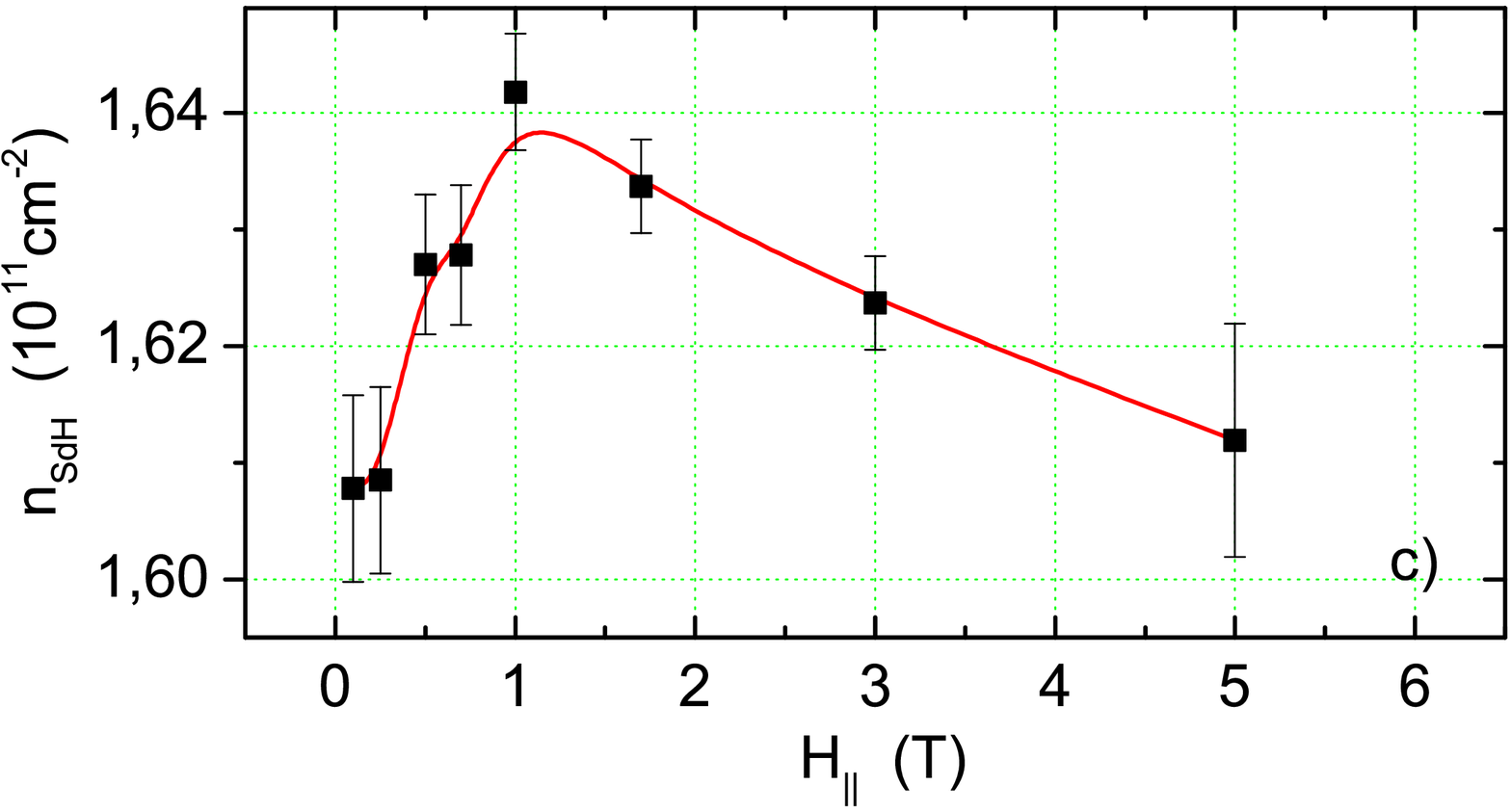}
\includegraphics[width=195pt]{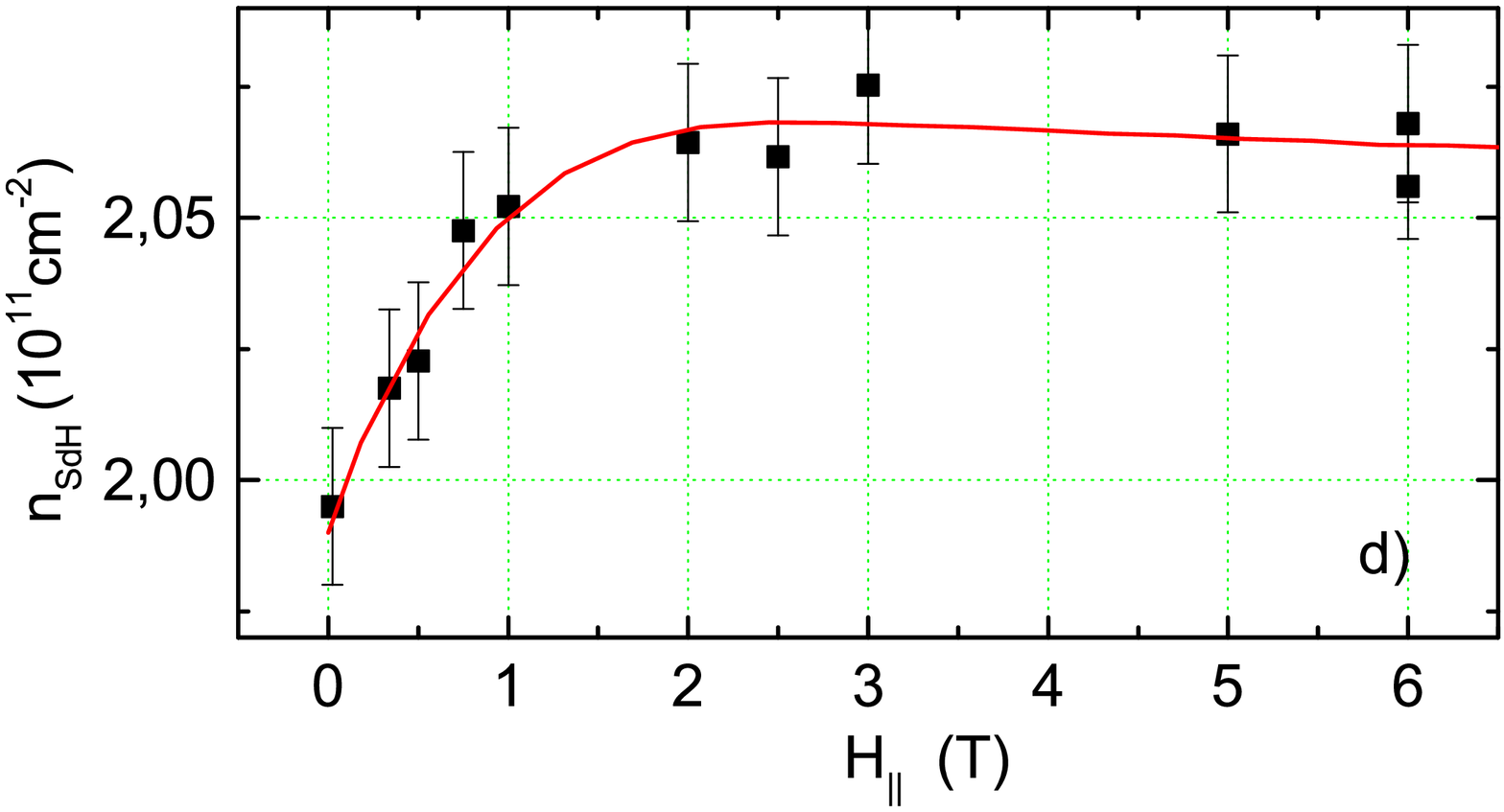}
\includegraphics[width=195pt]{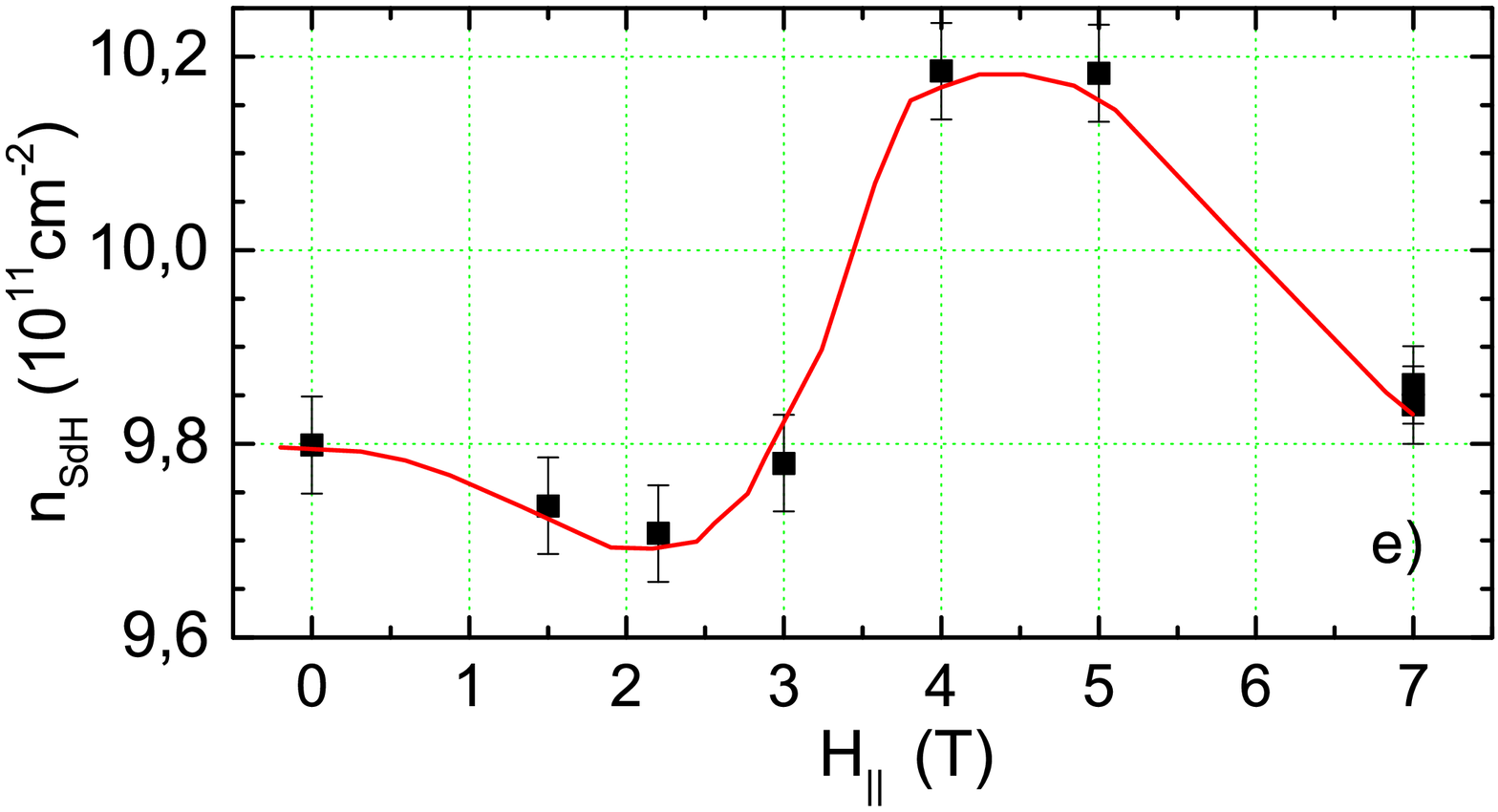}
\caption{Density  variations as a function of the in-plane magnetic field $H_\parallel$ for the zero-field densities  $n_{\rm SdH}(0)$: (a) 1.153, (b) $1.265$, (c) $1.64$, (d) $2.0$, and (e) $9.8 \times 10^{11}$cm$^{-2}$.
}
\label{fig4}
\end{figure}

 The  $\delta\chi^*/\chi(0)$ variations become smaller with increasing density:
e.g. at    $n= 6.16\times 10^{11}$cm$^{-2}$
 the drop  becomes a factor of 4 weaker than that for $0.99\times 10^{11}$cm$^{-2}$ [compare Fig.~\ref{fig3}a  and  Fig.~\ref{fig3}c].
For even a higher density $10\times 10^{11}$cm$^{-2}$,  $\chi^*(H)$ continues decreasing with field, and its overall change does not exceed $\approx 6\%$, [see Figs.~\ref{fig3}c and Fig.~\ref{fig3}d]. The minimum $\chi^*(H)$  shifts toward stronger fields, and the $V$-shape is replaced with a more complex non-monotonic dependence.
Figure \ref{fig4}  shows that the concomitant $n(H)$-dependence also changes from $\Lambda$-shaped to  a more complex one, and the initial rise of $n(H)$ is replaced with the $n(H)$ decrease in weak fields.
Though the shapes of the dependences $n(H)$ and  $\chi^*(H)$ become more complex, the correlations between them persist (Fig.~\ref{fig5}).  This case is discussed below.

\section{Conclusions drawn from the data}

(1) The observed small (2\%)
density variation (Fig.~2b)  cannot be the driving force
behind the $\sim 13\%$  variations of $\chi^*(H_\parallel)$ [Fig.~2a].
Indeed, one might expect $(\partial \ln \chi^*/\partial H)$ to be only  $\approx 0.6\%$ per Tesla for $n=1.5\times 10^{11}$cm$^{-2}$, estimated using the measured $\partial \chi^*/\partial n$ value from Ref.~\cite{gm}. Thus, the observed  $\delta n_{\rm SdH}(H)$ should be considered as a concomitant effect
rather than the main reason for $\chi^*(H)$ variation.

(2) The spin susceptibility variations $\delta\chi^*(H)$  measured from  SdH oscillations are relevant to
the mobile carriers. This data also correlate with with thermodynamic magnetization data [see Fig.~\ref{fig2}c], which is determined mainly by the large-spin collective localized states \cite{teneh_PRL_2012}.
Thus, we believe that the observed changes in the properties of extended states are caused by the magnetization changes of the localized states  and by the subsequent carriers redistribution between the two subsystems.

(3) The energy of the localized states
must be located in the close vicinity of the Fermi energy  to
allow  for the  carrier exchange  at ultralow temperatures between two electronic phases.
We refer to these states as ``fast'' localized states.  No temperature dependence of $\delta n_{\rm SdH}$ was observed within the range $0.1-0.5$\,K, therefore we believe that the carrier redistribution occurs elastically, via tunneling.  The corresponding energy diagram for  the two states is schematically shown in Fig.~\ref{fig7}c.  Note, that this picture is different from the conventional model of the disorder-localized single-particle  states  in the tail of the conduction band \cite{ando, gold_JPCM_2002, vitkalov_PRB_2002}.

(4) The  considered  ``fast'' localized states
are of a 2D nature and, hence, are a part of the 2D electron system rather than 3D interface states,
because both  $n_{\rm SdH}$ and $\chi^*$
changes are observed only in the in-plane field.

\section{Data overview}
Our measurements have been performed at a fixed gate voltage $V_g$, whereas
$H_\parallel$ and $T$ have been varied. Under this condition the total charge is conserved.
The total charge in the multi-component system
includes both the extended (``mobile'') and localized electron states.  Only the ``mobile'' carriers contribute to the SdH oscillations; their local density  is  experimentally found  from the oscillations frequency.

\begin{figure}
\includegraphics[width=240pt]{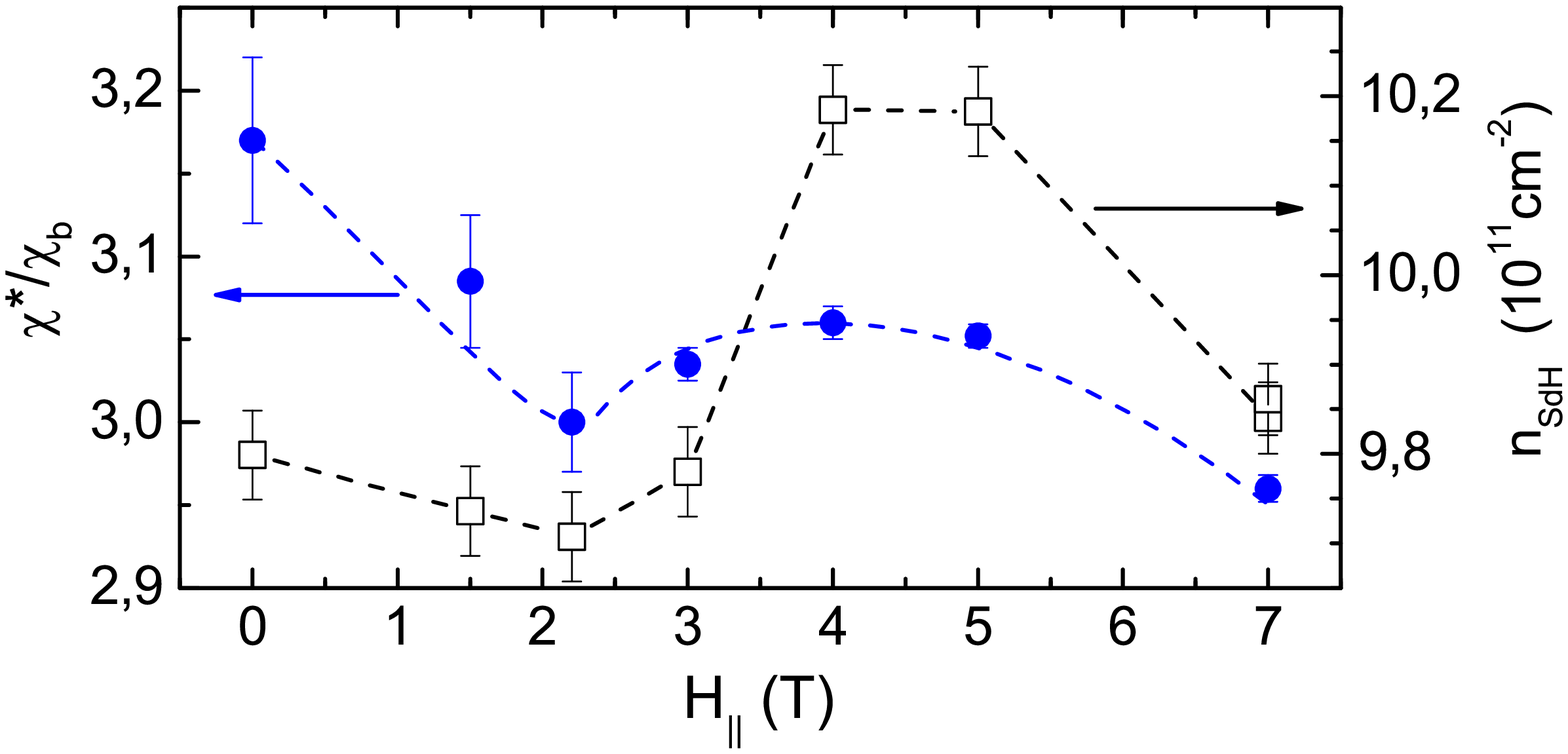}
\caption{Comparison of the magnetic field dependences of $\chi^*(H)/\chi_b$  and $n(H)$ at high carrier density $n\approx 10\times 10^{11}$cm$^{-2}$.   Sample Si3-10.
}
\label{fig5}
\end{figure}

The  localized states include ``fast''  and ``slow''  localized states. The former states are capable of recharging and reaching equilibrium with the extended states at the ms-time scale
after changing the  gate voltage, magnetic field and temperature
\cite{teneh_PRL_2012, tupikov_NatCom_2015}.
We associate the ``fast'' localized states with collective ``spin droplets''
possessing large spins. Such a conclusion, consistent with that deduced in Ref.~\cite{teneh_PRL_2012}, is  based  on (i) the observation of the large thermodynamic spin
magnetization exceeding the Bohr magneton \cite{teneh_PRL_2012},
fast increase of the magnetization energy with field,
and (ii)  the low value of the characteristic field of the magnetization saturation  $H\sim 0.25k_BT/\mu_B\approx 1$\,T [see Fig.~\ref{fig2}c].
The magnetization resembles that of free spins $M \propto \tanh(\mu_B H/k_B T)$,
but saturates in the field $H_\parallel \approx (0.8 - 1)$T, which is four times weaker than that anticipated for free spin - 1/2 electrons; this observation points at a large total spin $S\approx 2$ of the collective state \cite{teneh_PRL_2012}.

In contrast, the ``slow'' single-particle localized states (SPL), which are positioned deeply  below the Fermi energy,  in the tail of the conduction band, do not recharge within the time of measurements; they do not participate in  low-temperature transport and equilibrium thermodynamics.
Their presence may be revealed,
e.g. by polarizing the system with  the in-plane field \cite{gold_JPCM_2002, vitkalov_PRB_2002}.
 Since the SPL states are not recharging in the described measurements, we will not discuss them below. Correspondingly, we assume,  $\delta N_{\rm mob} + \delta N_{\rm loc}=0$ for the two-component system, where the  electron densities $\delta N_i$ in the two phases are functions of the temperature and magnetic field. We  show below that the  minimal phenomenological model involving these two components is capable of explaining our results qualitatively.

\section{Model}
In Appendices A and B to this paper we consider the spin susceptibility variations within the theory of interaction quantum corrections and the conventional thermodynamics of the single-phase state. We show that these effects cannot explain the reported experimental data. Moreover, the  observed magnetic field variation of the mobile carrier density under fixed  total charge in the gated structure is a clear indication of the presence of two phases in the studied electron system. For this reason we consider below the phenomenology of the phase separated two-phase state.

We conclude that the predicted magnetic field dependences due to interaction corrections and spin polarization are irrelevant (to the first approximation) to the observed sharp changes of $\chi^*(H)$.

\subsection{Thermodynamics of the two-phase state}
In the  phenomenological model proposed below, for
simplicity, we neglected both monotonic quantum corrections
and the dependence of $\chi^*$ on the degree of spin polarization.

Let us denote  $A_{FL}\equiv A_1$ - the fraction of the total area occupied by the FL states and $A_{\rm SD}=A_2$ -- the fraction of the total area occupied by the  collective localized states (spin droplets, SD).
$A_{SD}+A_{Fl}=1$, if one ignores the presence of the single-particle localized (SPL) states
(this can be justified since they do not recharge and do not thermalize during measurements).
 As long as the 2D system is conductive, the percolating-type arguments suggest that the FL phase is the majority phase and occupies more than 50\% of the sample area,  i.e. $A_{SD} < A_{Fl}$.

The number of electrons  per unit area in each phase  is  $N_1\equiv N_{\rm FL}$ and $N_2\equiv N_{\rm loc}$, and  $n_{\rm FL}$ and
$n_{\rm SD}$ correspond to  the local  densities of the states  in the regions occupied by FL- and SD- phases.  On the spatial scale  shorter than the distance between the gate and the 2D-layer  ($\approx 200$nm) the local densities in these two phases may only insignificantly  differ from each other and from the average density, both being dependent on external parameters. The areal density of electrons (i.e. per unit area of the overall 2D system) in each phase is $N_i=n_i A_i$.  Then the total charge of the 2D system
\be
Q/e\equiv N= N_{\rm 1} +N_{\rm 2}=
n_{\rm FL} A_{\rm FL} +n_{\rm SD} A_{\rm SD}.
\ee

The gate voltage controls $N$ rather than $n_i$ when the total 2D system is recharged.
 Correspondingly, for the two-component system at a fixed gate voltage $V_g$, $\delta N_{\rm FL} + \delta N_{\rm loc}=0$, where $\delta N_i$ are functions of temperature and field.
 The thermodynamic magnetization measurements  \cite{teneh_PRL_2012}
 show that the individual spin droplet size is independent of $N$,  $T$ and $H$.
 Hence, the local density $n_{\rm SD}$ also remains constant, whereas $n_{\rm FL}$ and $A_i$ may vary with
 $N$,   $T$ and $H$.
Based on the electrostatic arguments, we assume that $N$ depends only on $V_g$ , being independent of $H$ and $T$.

There is a fundamental difference
between the parameters probed by the transport and thermodynamic measurements. From frequency of the SdH oscillations one  determines (i) the local density of electrons participating in the cyclotron motion $n_{\rm SdH}\equiv n_{\rm FL}= N_1/A_1 $ and (ii) $\chi^*$ - the spin susceptibility of mobile electrons. On the other hand, both phases contribute to the thermodynamic magnetization, whose measurements provide $d\mu/dH= - dM/dN$ and $M =\int(dM/dN)dN$ ($M$ is roughly  proportional to $N_2= n_{\rm SD}\times A_{\rm 2}$ because the magnetization of the FL state is significantly smaller at  low and intermediate carrier densities). Similarly, the capacitive-type measurements performed at low frequencies, $\ll 10^{11}$\,Hz, probe all carriers, including SD and mobile FL states.

\subsubsection{On the origin and structure of the localized states}
In the absence of direct microscopic data on the spatial extension of the spin droplets (SD) and their energy spectrum, we can only conjecture on the SD origin. The  size of the spin droplet was estimated in Ref.~\cite{teneh_PRL_2012} as $\sqrt{2S/n}$, that is $\sim (40-100)$\,nm for the total spin $S=2$ and density $n= (0.5-2)\times 10^{11}$cm$^{-2}$. This size is comparable with the gate oxide thickness of 200\,nm; at greater distances the potential fluctuations are screened by the gate electrode. This comparison suggests that the spin droplets might originate due to the potential fluctuations at the Si-SiO$_2$ interface. However, without taking into account the intra-droplet e-e interactions, it would be difficult to explain why the total spin is so large and why the SD size and spin remain unchanged over a wide density range.

In our view, the most likely reason for the emergence of SD states is the Stoner-type instability that occurs locally in the most depleted regions. The interaction parameter $r_s$ is the largest in these regions, similar to the case of quantum dots discussed in Refs.~\cite{eisenberg_PRB_1999, brouwer_PRB_1999, kurland_PRB_2000, narozhny_PRB_2000, rogge_PRL_2010}.
An interesting issue is the ``magic'' total spin $S$ of SDs that is independent of temperature and the average density. The experimentally estimated $S=2$  \cite{teneh_PRL_2012}  indicates  that there are at least $n \geq 4$ electrons per SD. One might associate $n=4$ with the four-fold valley and spin degeneracy of the electrons at the (100)Si surface. We think that the valley splitting  at zero perpendicular field and valley degeneracy are irrelevant since for the samples studied $\Delta_v$ (0.4\,K for Si6-14)  is less than the temperature of measurements in Ref.~\cite{teneh_PRL_2012}. The Zeeman splitting $\Delta_Z$  for spins 1/2 is also irrelevant, since the measurements in Ref.~\cite{teneh_PRL_2012} as well as in this paper have been performed at $\Delta_Z < k_B T$; under such conditions all spins and valleys are mixed. We therefore believe that the total spin and the number of electrons in a droplet are set by the intra-droplet many-body interactions. Note that the maximum spin corresponds to a maximally antisymmetric coordinate wave function, which reduces Coulomb repulsion in restricted geometry.

Our observation that the SD states can recharge and quickly thermalize  with the FL-states indicates that the energy band of the SD states  must be located in the vicinity of the Fermi level. This enables the $T$-independent elastic carrier exchange between the two bands  via tunneling rather than the temperature-induced activation.
Since the collective states are energetically more favorable in the restricted area, the uppermost energy level of the SD state is likely to be  located somewhat lower than $E_F$, by the binding energy $\Delta$ of the collective state.
%, as schematically shown in the inset of Fig.~5d-SM.

In analogy with the conventional quantum dots and self-organized potential traps  \cite{tripathi_PRB_2006},
here the confining potential of the SD state should be surrounded by a barrier preventing the dot from filling with  electrons from the neighboring FL phase (Fig.~\ref{fig6}b,c).  We believe that the SD states emerge near the maxima of the bare fluctuating potential, where a local collective (Stoner or spin polarized Wigner) state emerges in the most depleted areas. This conclusion seems counterintuitive, since conventional single-particle localized states are located {\em at the bottom of the potential landscape} and therefore fall into the band tail. The ground state energy of SD becomes lower by the binding energy $\Delta$  than $E_F$ of the surrounding Fermi sea (Fig.~\ref{fig6}c). The uppermost level of the collective state then goes down, below the peak of the potential hill (like a ``volcano crater'') leaving the surrounding potential barrier almost intact. Leaving aside the origin of the emerged ``crater'' (i.e. the SD energy level), the formation of the surrounding barrier is governed further by the conventional screening similar to that in Ref.~\cite{tripathi_PRB_2006}.

The tunneling resistance $R$ of the barrier and the SD size-dependent capacitance provide the characteristic $RC$-time required for establishing equilibrium when $T$, $H$ or $N$ are varied. Note that we ignore the strongly  localized states positioned well below the Fermi energy, in the band tail, since they don't thermalize within the time of measurements. Accounting for these deep traps simply changes the overall density  $N$  by an offset that is temperature- and field-independent
\cite{ando}.

\subsubsection{Magnetic field variations of $M$ and $n$}
The free energy of a multicomponent system is \cite{landau}
$F = \sum_i\mu_i N_i$,
 where $\mu_{1,2}=\partial F/\partial N_{1,2}$ are the electrochemical potentials of the extended (FL) and localized (SD) components.
In equilibrium $\mu_1(H,T)$ and $\mu_2(T,H)$ are equal, however they  may  depend differently  (and, in fact, do depend) on $H$ and $T$.

\begin{figure}
\includegraphics[width=180pt]{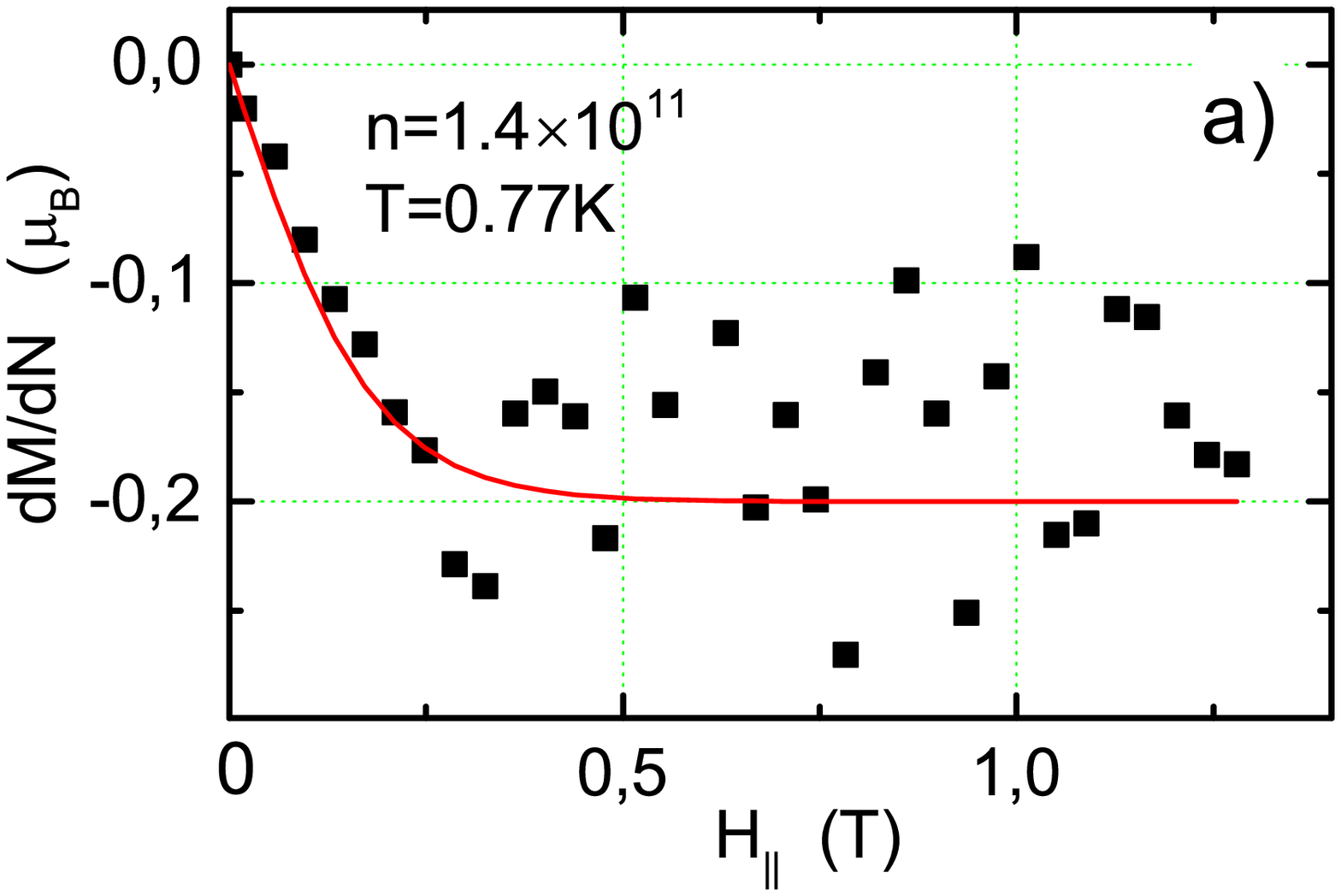}\\
\hspace{0.25in}\includegraphics[width=160pt]{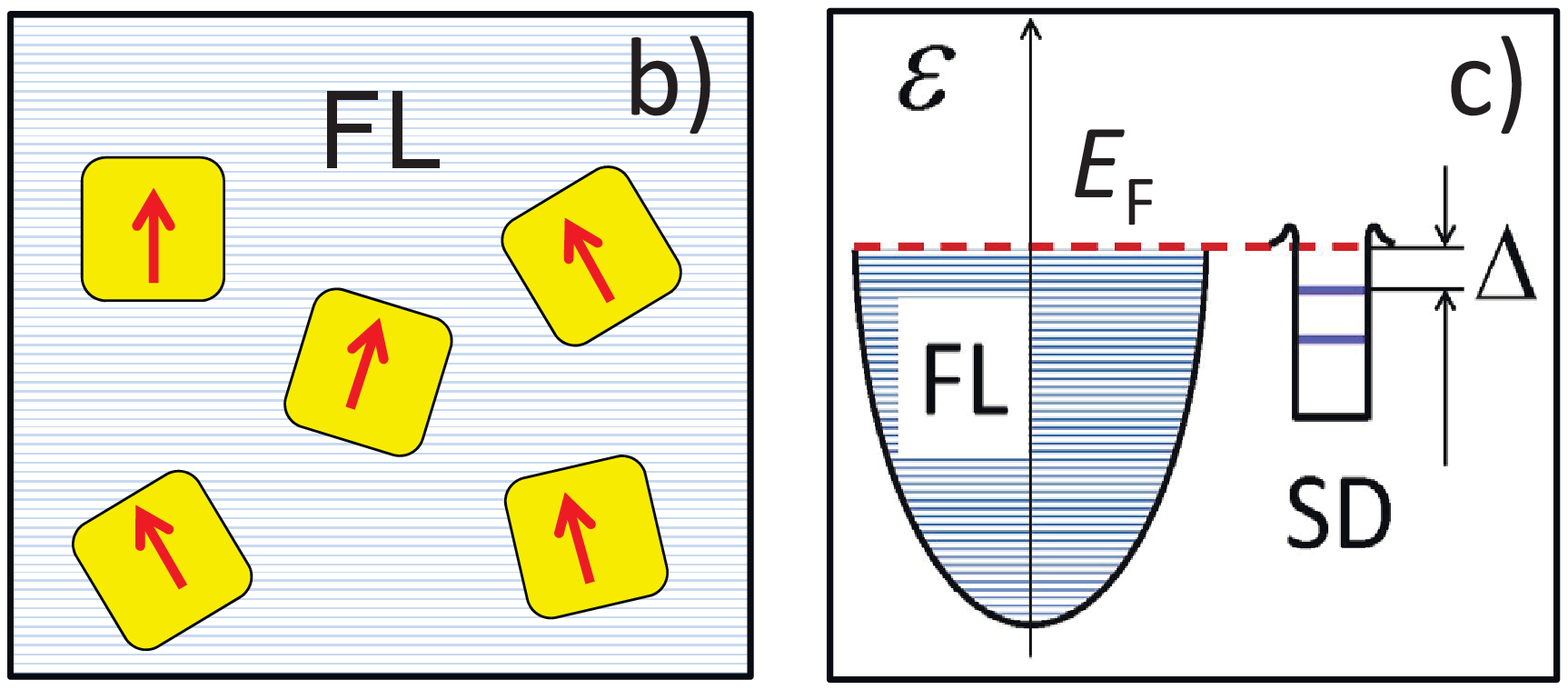}
\caption{(a) $dM/dN$ vs $H_\parallel$. Symbols are the data from
Ref.~\cite{pudalov_PRB_2018} for $n=1.4\times 10^{11}$cm$^{-2}$.
 (b,c) Schematic spatial arrangement of the two-phase state and  the energy band diagram of the two-phase system.
}
\label{fig6}
\end{figure}

Taking into account the result of Ref.~\cite{teneh_PRL_2012}
that the individual size of each SD (or quantum dot) does not change with field and temperature, the  changes $N_2(B)$ may be induced only by variation of the number of SD states per unit area, i.e. by $A_2$. It is convenient to split the total free energy per unit area into three terms:

\bea
F= F_\mu +F_{el}+F_H = \left[\mu_1 N_1 +
\mu_2 N_2\right]  - \\
\left[\frac{e^2N_1^2}{2C_1} + \sum_j \frac{Q_{SD}^2}{2C_{SD}}\right] - \left[\frac{\chi_1 H^2}{2} + M_2 H\right].
\label{eq:3parts}
\eea
where $\mu_1=\mu_2$ are the chemical potentials of each phase,
% at $H=0$, which are equal in equilibrium,
$C_1 \approx (A_1)^{1/2}$ and
$C_2 \approx (A_2)^{1/2}$ -- are the capacitances of the  FL-state and SD-state per unit area,  $M_2(H)$ - the magnetization of the SD state per unit area, $\chi^* (H)$- the spin susceptibility of the FL state per unit area, and the sum is over all SD-states per unit area.
The total electrostatic energy  equals $N^2e^2/2\sqrt{A} = N^2e^2/2$ and is independent of $H$ and $T$.
In the free energy we took into account the electrostatic energy, magnetization energy of the mobile
FL states and of the localized SD states (see Fig.~\ref{fig6}b),  and, for simplicity,  neglected  the mixing energy at the boundaries \cite{castellani_PRB_2001}.

In equilibrium,  we require
\cite{landau}
\be
\delta F= \delta F_\mu +\delta F_H +\delta F_{el}=0,
\label{Eq:equilibrium}
\ee
where variation of $F(H,N_i)$ is taken with respect  to all relevant  variables,  $H_\parallel$, and $N_i$. In what follows we consider only the case of a fixed temperature, since the available data
\cite{pudalov_PRB_2018, teneh_PRL_2012}
on the $T$- dependences of $M$, $\chi^*$, and $n_i$ have  insufficient accuracy.

After regrouping terms in Eq.~(\ref{Eq:equilibrium}),  dividing them by $\delta H$, and taking into account that
 $\delta N_1=-\delta N_2$, Eq.~(\ref{Eq:equilibrium}) reduces to
\be
\frac{\partial N_1}{\partial H}\left[\frac{N_1}{\tilde{D}_1}-  \frac{N_2}{\tilde{D_2}} +\frac{\partial M_2}{\partial N_2}H \right] =  -\frac{\partial \mu_2}{\partial H}N_2 - \frac{\partial \mu_2}{\partial H}H +M_2,
\label{Eq:balance}
\ee
where $\tilde{D}_i=\partial N_i/\partial\mu_i$, and  $D_i = \partial n_i/\partial \mu_i$ is the thermodynamic density of states for the $i-$th component.
\be
\tilde{D}_i=\frac{\partial N_i}{\partial \mu_i} = A_i\frac{\partial n_i}{\partial \mu_i} + n_i \frac{\partial A_i}{\partial \mu_i}= A_i D_i +n_i \frac{\partial A_i}{\partial \mu_i}.
\ee

The third term in the square brackets of Eq.~(\ref{Eq:balance}) may be neglected since it is by a factor of $\sim 70$ smaller than the first one. We also neglected the terms  $(\chi H)$ and $(\partial \chi/\partial H)H^2$ in $\delta F_H$ because of their smallness.
Substituting
$(\partial \mu/\partial H)=-(\partial M/\partial N)$  we obtain from Eq.~(\ref{Eq:balance})
\be
\frac{\partial N_1}{\partial H}\left[\frac{N_1}{\tilde{D}_1}-  \frac{N_2}{\tilde{D}_2} \right] =  \frac{\partial M_2}{\partial N_2}N_2 + \frac{\partial M_2}{\partial H}H +M_2,
\label{Eq:5}
\ee
and after  integrating both parts with respect to $H$:
\begin{align}
N_1(H)&\left[\frac{N_1}{\tilde{D}_1}-  \frac{N_2}{\tilde{D_2}}\right] = \nonumber \\
\int dH &
\left[ \frac{\partial M_2(H)}{\partial N} \frac{\partial N}{\partial N_2} N_2
+ \frac{\partial M_2(H)}{\partial H}H +M_2(H)\right]
\label{int_equality}
\end{align}

Equation (\ref{int_equality}) relates changes in magnetization $M_2(H)$ of the
SD localized states with the density variations
$N_1(H)$ in the majority FL phase.  In this equation $M_2(H)$, $\partial M_2/\partial H$, and $N_1(H)$  have been determined experimentally (see Fig.~\ref{fig2}), and  $\partial M_2/\partial N$
may be found from Ref.~\cite{teneh_PRL_2012} for a given density.
The nominal carrier density $N_1\propto n_{\rm SdH}$ is known from the SdH measurements, and $D_1$, in principle, may be calculated from the known renormalized effective mass $m^*$ \cite{gm}.

\subsection{Comparison  of the model with experimental data}
\begin{figure}
\includegraphics[width=230pt]{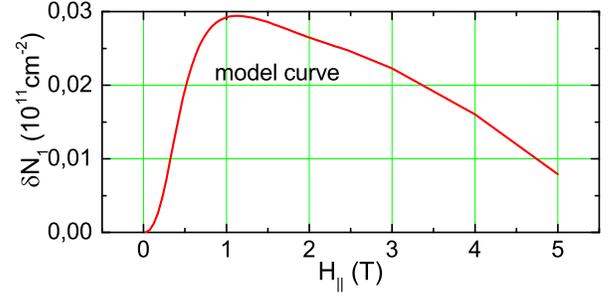}\\
\caption{Model curve $\delta N_1(H_\parallel)$ calculated from experimental data as described in the text
}
\label{model}
\end{figure}

 The dependence   $n_{\rm FL}(H) \propto N_1(H)$  calculated  using Eq.~(\ref{int_equality})
 for the representative density $1.4\times 10^{11}$cm$^{-2}$
  is shown in Fig.~\ref{model}.
It  can be compared with direct experimental data of Fig.~\ref{fig2}c.

Although  the $M_2(H_\parallel)$  and $n_{\rm SdH}(H_\parallel)$ data are available at slightly different densities,  and the model has several simplifications, the calculated $\delta N_1(H)$ dependence captures the main features of the experimental data:
the density of mobile electrons grows  with field, reaches a maximum at approximately the same field of 1T, and than slowly decreases. We conclude that there is
a good qualitative  agreement with the experimentally measured $n_{\rm FL} (H)$.
On the r.h.s. of Eq.~(\ref{int_equality}) the second and third terms  are positive and monotonically grow with field. Only the first term is negative and, thus, results in  the  maximum and subsequent decrease in
$N(H)$. This  term is  set to $(3.36\pm 0.1)\times 10^{11}\mu_B$/cm$^2$ in order to satisfy Eq.~(\ref{int_equality}) and fit the observed decrease of $N(H)$.

 For high densities,
we believe that the complex shapes of the $\chi^*(H)$ and $n(H)$ dependences (Figs.~\ref{fig3}e and \ref{fig4}d) reflect the competition of several effects. One possibility is that the energy of the spin magnetization of the FL states, $\chi^* H^2/2$, (which we assumed to be negligibly small at low and intermediate density,  and omitted for simplicity in our model)  becomes comparable with (and may exceed) the magnetization energy of the SD states, $M_2H$. This is because the number of SD states vanishes as overall density increases
\cite{teneh_PRL_2012, morgun_PRB_2016}.
Indeed,  if we ignore the SD states magnetization for high densities,  the monotonically growing with field FL magnetization energy $\chi^* H^2$ (in contrast to the sharply saturating $M_2H$) in  Eq.~(6) would produce monotonic decrease of $n(H)$.
The observed initial $n(H)$ decrease with field (Fig.~\ref{fig3}e) is consistent with such interpretation and indicates transfer of electrons from FL to the SD states. However,
 the accuracy and completeness of our data is insufficient to quantitatively treat these effects at high densities.

\section{Discussion}
1. The similarity of the modeled and  measured
$n_{\rm SdH}(H)$ data justifies our approach.
In the proposed scenario, the magnetization of the SD states causes changes in the free energy, which, in turn, cause redistribution of the carriers between the extended and localized states. The changes in the free energy also affect the magnetic energy of the delocalized states (detected experimentally as $\delta\chi^*(H)$). These changes were neglected in our model, since they are determined by a difference of two much larger terms in $\delta F_H$  and the accuracy of our data is insufficient to calculate them.
With  rising density,
 the amount of the SD states diminishes \cite{morgun_PRB_2016}, and the contribution of the delocalized FL states to magnetization energy  becomes dominant. Apparently, this is the reason  for the
 evolution of shape of the experimentally determined  $n_{\rm SdH}(H)$ and $\chi^*(H)$ dependences at the highest carrier density $10\times 10^{11}$cm$^{-2}$ (see Fig.~\ref{fig5}.

2. It is worth noting that for a 2D FL  system, due to the electron-electron interaction corrections in the diffusive regime $T\tau \ll 1$, the spin susceptibility is expected to vary smoothly and insignificantly $\propto -\ln(g\mu_B H/2\pi k_B T)$ in fields $g\mu_B H_\parallel \ll k_B T$  \cite{pudalov_PRB_2018, altshuler_JETPL_1982} (see also Eqs.~(\ref{eq:lowB}) and (\ref{eq:strongerB}) in Appendix A). With approach to the field of complete spin polarization of the 2D FL system [$H_p \sim 2E_F/g\mu_B \sim 10 - 20$T for the relevant densities $(1- 2)\times 10^{11}$cm$^{-2}$], the spin susceptibility is predicted to strongly increase  \cite{zhang_PRL_2006}. Anyhow, $\chi^*$ in a single-phase 2D FL system is not expected to exhibit strong variations in  weak fields of the order of $k_BT/\mu_B$.

3. Variation of the carrier concentration in 2D gated systems \cite{pudalov_JETPL_1984, pudalov_JETPL_1986}  in perpendicular field is a well-known effect.
 The variations $\delta N(H)$ at a constant gate voltage are commonly treated  within the framework of the single-phase picture and related to the chemical potential jumps $\delta\mu(H)$ between the Landau levels in the spectrum of the FL state.
 Alternatively, in thermodynamic magnetization measurements in the weak in-plane field \cite{teneh_PRL_2012, reznikov_JETPL_2010},  recharging of the gated 2D structure was predominantly caused by the SD states, whereas the
 transport response of the  mobile FL states was not measured. Therefore, it is possible to treat the results within single-phase model and almost ignore the FL states.

  In the current paper we measured and analyzed both the transport and thermodynamic data, and found that the conventional single-phase approach is not capable to explain the two sets of data even  qualitatively.
This is because in the parallel field the magnetization of the SD subsystem exceeds (or at least, is comparable with) the Pauli magnetization of  mobile electrons.

4. The carrier redistribution between the localized and extended states is common for many types of field-effect semiconductor-insulator structures \cite{sirringhaus_AdvMat_2009, lee_PRB_2010}. This effect typically ``freezes out'' at lowering temperatures, because it requires temperature activation from the tail states to the Fermi level. Here we reported the effect of  redistribution that persists down to ultralow temperatures.  We associate  this temperature independent effect with elastic tunneling between the states of different nature but with the same energy.

5. The carrier redistribution between two phases in the 2D system  is not easy to determine by other techniques. For example, the capacitance measurements taken at frequencies
$0-10^6$\,Hz (1\,nF, 1\,kOhm/$\Box$)
probe the total charge density that includes both SD and mobile states.  To separate the SD and FL states, the capacitance measurements should be done at frequency of ~$10^{10}-10^{12}$Hz, inaccessible for the gated structure.
We already noted that the FL density deduced from SdH oscillations in the phase-separated system is determined by the local density in the FL lakes (where the carriers possess the highest relaxation time), rather than by the total density; this picture holds until the delocalized states (FL lakes) percolate. For the gated 2D system, the difference of the local density within the FL lakes and within the SD areas is negligibly small because of the associated electrostatic energy and neutrality requirement.

The difference between the overall charge in the 2D system (determined from recharging measurements) and the mobile FL carrier density (local charge density) deduced from SdH/QHE  was measured in Ref.~\cite{pudalov_JETPL_1984} and found to be less than 2\%, i.e. within the experimental uncertainty.  Note, that the Hall voltage measurements also cannot shed light on the mobile and SD carrier density, since the Hall voltage becomes irrelevant to the carrier density at the verge of the localization transition (the so called ``Hall insulator'') \cite{pudalov_JETPL_1993}.

6. Though the $\chi^*(H_\parallel)$ changes for mobile carriers were not calculated in our model because of the smallness of the corresponding magnetic energy changes, the changes should occur due to the following thermodynamic arguments. The microscopic mechanism behind these changes is as follows: the carriers in the SD states have their spins aligned ferromagnetically along the field, and when they tunnel elastically into the Fermi sea, they should join the spin-minority subband. As a result, the spin polarization degree $(n\uparrow - n\downarrow)/n$ measured from the SdH oscillations beating diminishes;  the spin susceptibility reported here has been found from this parameter. Further slow restoring of $\chi^*$ with $H_\parallel$-field presumably reflects the spatially averaged density of states for the SD's. Qualitatively, we conjecture that the average width of the uppermost energy level in the SD states amounts to $\sim 4\,$K,  as estimated from the corresponding restoration field of 3\,T  in Fig.~2.

7. The observed $\chi^*(H_\parallel)$ variations (Figs.~1, and 2) indicate that a widely used approach for determining the spin susceptibility from the
monotonic magnetoresistance  measurements in  parallel magnetic fields \cite{SKD_MR, vitkalov_MR}  might result in significant errors. Indeed, the
$\langle g^*m^*(H_\parallel)\rangle$ averaged over a wide field range (from 0 to the spin polarization field $H_p$)
is about 10\% smaller than the zero-field value $g^*m^*(H\rightarrow 0)$.
If the monotonic magnetoresistance is measured in weaker fields $H \leq H_p$,  the
underestimation of $\chi^*$ caused by finite $H_\parallel$ may be even greater (e.g., it may reach 15\% for the degree of spin polarization $\approx 0.05-0.1$).

\section{Conclusions}
We have  observed and explored  an unexpected sharp field dependences
 of the FL  spin susceptibility  $\chi^*(H_\parallel)$
 and the density  $n_{\rm SdH}(H)$ of the mobile electrons  in the regime of strong inter-electron interactions
 ($r_s= 7.3 - 3.4$). The two effects correlate well with each other and with the thermodynamic magnetization of the localized SD states. We suggested a simple phenomenological
 two-phase model that links the  changes in the density of the mobile electrons to the magnetization of the collective localized states which are spatially separated from the extended FL states. The qualitative agreement of the model with experimental data
 suggests that the  variations of $n_{\rm SdH}(H)$ and $\chi^*(H)$ with the in-plane field are caused by magnetization of the minority phase of collective localized states. Thus, our results
  provide the solid evidence for the  phase separation in the interacting 2D electron system even at relatively high carrier densities, deeply in the ``metallic'' regime of high conductivity [$\sigma = (3 - 80)\times(e^2/h)$] \cite{gmax}.

 Our results also explain a long-standing  disagreement between the experimentally measured values of the spin-susceptibility and $g^*$-factor obtained in weak and strong magnetic fields.  Though  the presented empiric phenomenological model qualitatively explains the data at intermediate densities,  for the quantitative analysis a microscopic theoretical consideration is required that would take into account the energy spectrum of the SD states.

VP acknowledges fruitful discussion with A.~L.~Rakhmanov.
This work was supported by the RFBR, project No. 18-02-01013,
and fulfilled within the state assignment of the Ministry of Science and Higher Education
of the Russian Federation (Project ``Physics of high temperature superconductors and novel quantum materials''
No 0023-2019-0005).

\section{Appendix A:
Theoretically  predicted $\chi(H)$ dependence for the 2D Fermi liquid}
The Fermi liquid quantum correction to  $\chi(H)$ in the diffusive interaction regime
at  weak fields and not too low temperatures, $(1+\gamma_2)g\mu_B H_\parallel/(2 \pi k_BT) \ll 1$,  to the second order in $h$ is given  in Ref.~\cite{pudalov_PRB_2018}:

\begin{equation}
\frac{\delta  \chi^*(H_\parallel)}{\chi^*(0)}= -\frac{1}{\sigma_{xx}} \frac{\zeta(3)}{\pi^2}g_v^2[(1+\gamma_2)^3-1] h ^2,
\label{eq:lowB}
\end{equation}
where $h=g\mu_B H/2\pi k_BT$, $g_v$ - valley degeneracy, $\gamma_2=-F_0^\sigma/(1+F_0^\sigma)$, $F_0^\sigma$ - the FL coupling constant  in the particle-hole triplet channel,   and $\chi^*(0)= g_v \mu_B^2 m^*/[\pi(1+F_0^\sigma)]$ is the spin susceptibility of the FL at $H=0$, and $T=0$.
For stronger fields or lower temperatures $2 \pi k_B T \ll (1+\gamma_2) g\mu_B H_\parallel$,
\begin{equation}
\frac{\delta \chi(H_\parallel)}{\chi(0)} \approx  -\frac{1}{\sigma_{xx}}\frac{g_v^2\gamma_2}{\pi^2} \ln(g\mu_B H_\parallel \tau),
\label{eq:strongerB}
\end{equation}
which is consistent with the theory by Al'tshuler and Zyuzin
\cite{altshuler_JETPL_1982}.

Equations (1) and (2)  predict $\chi(H)$ to {\em  monotonically decrease} with $H_\parallel$, unlike  the experimentally observed  $V$-shape $\chi^*(H)$ dependence (see Fig.~1).
Within an alternative approach, by taking into account the spin polarization dependence of the exchange and kinetic energy, Zhang and Das Sarma found that   $\chi^*$ should {\em  monotonically grow} with $H_\parallel$ almost up to the field of complete spin polarization
\cite{zhang_PRL_2006}.

We stress that (i) the characteristic field (typically, $H_\parallel \sim 1$\,T - see Fig.~1) of the observed  $\chi^*(H)$ minimum is much weaker than the field of complete spin polarization of a pure 2D system $H_p=2E_F/g\mu_B \approx 10.88(N/10^{11})$[T], and  (ii)  there is no other characteristic field besides $H_p$  in a homogeneous single-phase system. We conclude that the predicted magnetic field dependences due to interaction corrections and spin-polarization are irrelevant (to the first approximation) to the observed sharp changes of $\chi^*(H)$.

\section{Appendix B: Thermodynamics of the single-phase state}
Consider the simplest single-phase picture that is traditionally applied to the capacitive type measurements with gated 2D systems. Namely, we consider the total charge density
\cite{pudalov_JETPL_1986, reznikov_JETPL_2010}:
\be
N= \frac{C_0}{e}\left(V_g-\frac{\mu_{\rm 2D}-\mu_g}{e} \right),
\ee
where $\mu_{\rm 2D}$, and $\mu_g$ are the electrochemical potentials of the electrons in the 2D layer and the gate (Al film), $V_g$ - is the gate voltage, and $C_0$ is the capacitance between the 2D layer and the gate. Here we neglect slight renormalization of the geometric capacitance by the electron compressibility
\cite{reznikov_JETPL_2010}.
Taking variations with in-plane magnetic field, we obtain:
\be
\frac{e^2}{C_0}\frac{\partial N}{\partial H}=+\frac{\partial M}{\partial N}
\ee

\begin{figure}
\includegraphics[width=180pt]{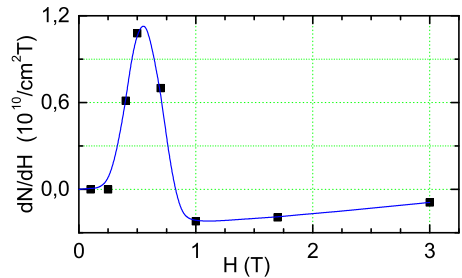}
\caption{Example of  the magnetic field dependence of  $dN(H)/dH$ calculated from Fig.~4c. Sample Si6-14. $N(0)=1.61\times 10^{11}$cm$^{-2}$
}
\label{fig8}
\end{figure}

Here, we used the Maxwell relation $\partial \mu_{\rm 2D}/\partial H = - \partial M/\partial N$
and disregarded the ``diamagnetic shift'', i.e. magnetic field dependence of $C_0$. The diamagnetic shift
contributes to $\partial M/\partial N$ less than $0.05\mu_B$ at low densities and further drops as  $N$ increases
 \cite{reznikov_JETPL_2010}.
This equation predicts direct proportionality between the magnetic field variation of  $dN/dH$  and  the magnetization per electron $\partial M(H)/\partial N$.
Both quantities have been experimentally measured:  $dN(H)/dH$ is shown in Fig.~\ref{fig8} and $\partial M(H)/\partial N$ -- in Fig.~\ref{fig6}a. Clearly, there is  little in common between the two dependences, and we  conclude that the single-phase picture is inadequate for explaining  the experimental results.

\end{document}